\documentclass[letterpaper]{article} 
\usepackage{aaai24}  
\usepackage{times}  
\usepackage{helvet}  
\usepackage{courier}  
\usepackage[hyphens]{url}  
\usepackage{graphicx} 
\usepackage{natbib}  
\usepackage{caption} 
\usepackage{algorithm}
\usepackage{algorithmic}
\usepackage{bm}
\usepackage{multirow}
\usepackage{amsmath}
\usepackage{amsfonts}
\usepackage{booktabs}
\usepackage{newfloat}
\usepackage{listings}
\DeclareCaptionStyle{ruled}{labelfont=normalfont,labelsep=colon,strut=off} 
\floatstyle{ruled}
\newfloat{listing}{tb}{lst}{}
\floatname{listing}{Listing}
\title{Cross-Gate MLP with Protein Complex Invariant Embedding Is a One-Shot Antibody Designer}
\author{
    Cheng Tan$^{1,2}$\equalcontrib, Zhangyang Gao$^{1,2}$\equalcontrib, Lirong Wu$^{1,2}$, Jun Xia$^{1,2}$, Jiangbin Zheng$^{1,2}$, Xihong Yang$^{3}$, Yue Liu$^{3}$, Bozhen Hu$^{1,2}$, Stan Z. Li$^{2}$\thanks{Corresponding author.}
}
\affiliations{
    \textsuperscript{\rm 1}Zhejiang University \\
    \textsuperscript{\rm 2}AI Lab, Research Center for Industries of the Future, Westlake University\\
    \textsuperscript{\rm 3}College of Computer, National University of Defense Technology\\
    tancheng@westlake.edu.cn
}

\usepackage{bibentry}

\begin{document}

\maketitle

\begin{abstract}
Antibodies are crucial proteins produced by the immune system in response to foreign substances or antigens. The specificity of an antibody is determined by its complementarity-determining regions (CDRs), which are located in the variable domains of the antibody chains and form the antigen-binding site. Previous studies have utilized complex techniques to generate CDRs, but they suffer from inadequate geometric modeling. Moreover, the common iterative refinement strategies lead to an inefficient inference. In this paper, we propose a \textit{simple yet effective} model that can co-design 1D sequences and 3D structures of CDRs in a one-shot manner. To achieve this, we decouple the antibody CDR design problem into two stages: (i) geometric modeling of protein complex structures and (ii) sequence-structure co-learning. We develop a novel macromolecular structure invariant embedding, typically for protein complexes, that captures both intra- and inter-component interactions among the backbone atoms, including C$\alpha$, N, C, and O atoms, to achieve comprehensive geometric modeling. Then, we introduce a simple cross-gate MLP for sequence-structure co-learning, allowing sequence and structure representations to implicitly refine each other. This enables our model to design desired sequences and structures in a one-shot manner. Extensive experiments are conducted to evaluate our results at both the sequence and structure levels, which demonstrate that our model achieves superior performance compared to the state-of-the-art antibody CDR design methods.
\end{abstract}

\section{Introduction}

Antibodies are essential proteins that the immune system makes to fight foreign substances, or antigens~\cite{raybould2019five,kong2023end,shi2022protein}. They have a Y-shape with two arms that can attach to specific antigens, such as bacteria. Once an antibody binds to an antigen, it marks the antigen for destruction by other cells in the immune system~\cite{basu2019recombinant}. The ability of antibodies to recognize and bind to antigens is crucial for the immune system's defense against infections and protection against future exposure to the same antigen~\cite{maynard2000antibody,akbar2022progress}. The complementarity-determining regions (CDRs) are parts of the variable domains of the antibody chains that vary and form the antigen-binding site that defines the specificity of the antibody~\cite{kuroda2012computer}. CDRs play a critical role in the recognition and binding of antigens by antibodies~\cite{tiller2015advances}. 

Early works on antibody design focus on generating the sequences of CDRs without the corresponding structures~\cite{saka2021antibody,alley2019unified,shin2021protein}, while a recent work~\cite{jiniterative} proposes a novel approach called RefineGNN, which enables the co-design of both the sequences and structures of antibody CDRs. DiffAb~\cite{luoantigen} proposes generating antibodies with high affinity to given antigen structures. MEAN~\cite{kong2023conditional} further involves the light chain context information as a conditional input to generate CDRs. However, as more conditional contexts are introduced, designed models struggle to adequately capture the complex interactions between complementarity-determining regions (CDRs) and context information. This is due to an insufficient geometric modeling approach, which relies solely on considering the C$\alpha$ atoms or orientations in each residue. 

\begin{figure}[ht]
\centering
\includegraphics[width=0.98\linewidth]{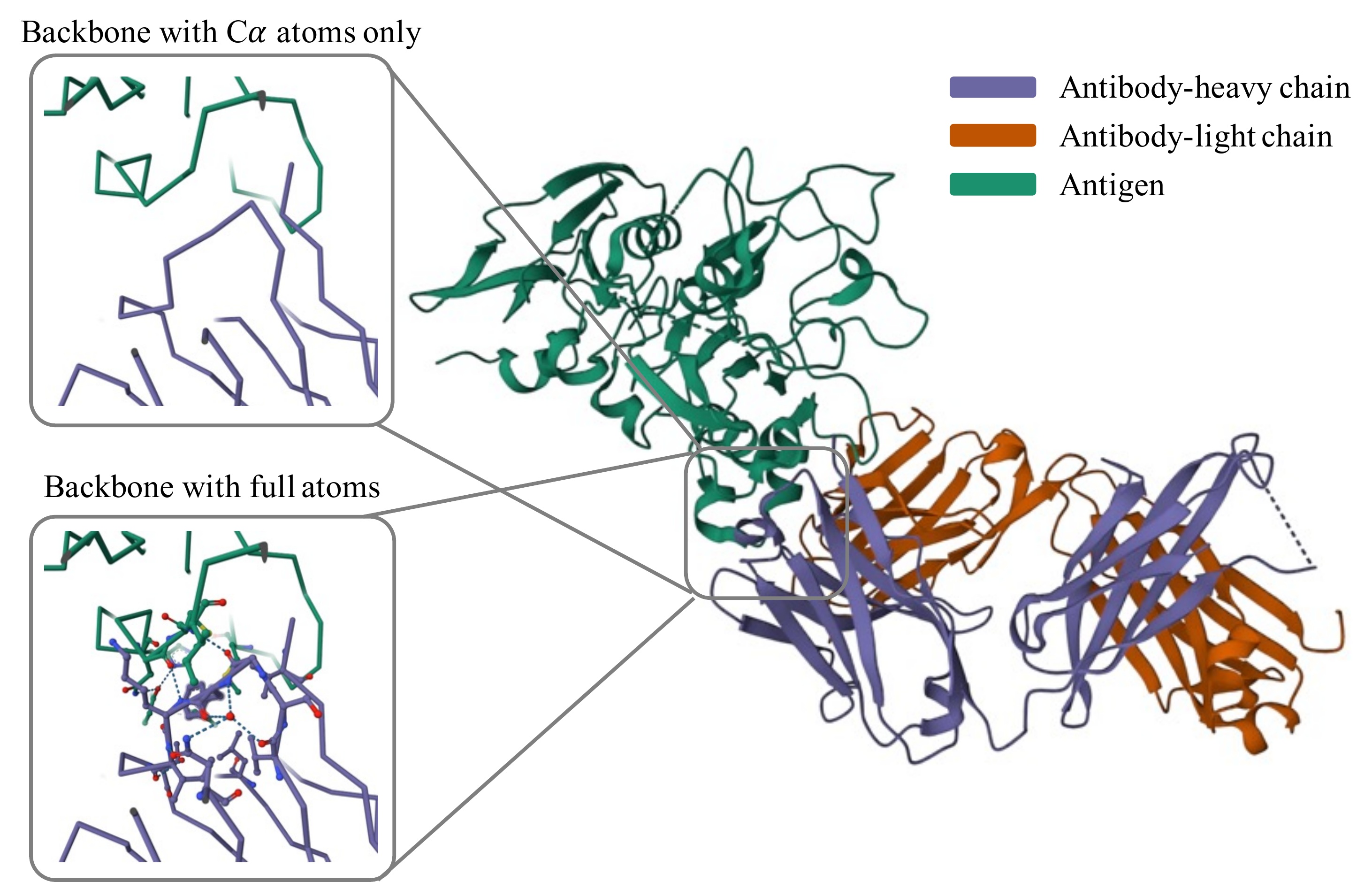}
\caption{The backbone comprised solely of C$\alpha$ atoms provides a reduced amount of information compared to the backbone consisting of all atoms.}
\label{fig:motivation}
\end{figure}

As demonstrated in Figure~\ref{fig:motivation}, the use of full backbone atoms provides a richer source of geometric information that is critical for the intricate interactions between components of antibody-antigen complexes. While some contemporary studies~\cite{jiniterative,luoantigen} have endeavored to integrate the orientations of amino acids, their capacity to furnish comprehensive information remains constrained without the utilization of full backbone atoms. Moreover, existing works rely on either iterative refinement~\cite{jiniterative,fu2022antibody,kong2023conditional} or diffusion sampling strategies~\cite{luoantigen} in the CDR decoding process, which leads to inefficient inference. 

To address the limitations of existing methods, we propose a novel antibody CDR design model that co-learns the 1D sequences and 3D structures using a protein complex invariant embedding network and can decode the CDRs in a one-shot manner. Specifically, we decouple the antibody CDR design into a two-stage process: (i) \textbf{geometric modeling of protein structures} and (ii) \textbf{sequence-structure co-learning}. For the comprehensive geometric modeling, the protein complex invariant embedding constructs intra-component relationships inside the same component and inter-component interactions between different components of antigen-antibody complexes with full backbone atoms. Our approach explicitly models complete atomic-level geometric dependencies, which include not only C$\alpha$ atoms but also N, C, and O atoms in the protein backbone. This enables our model to capture the intricate dependencies between the CDRs and the contexts. Then, we introduce a simple cross-gate MLP to implicitly refine the sequence and structure representations by each other in a co-learning manner. The model thus does not require any explicit iterative refinement strategies that are computationally expensive. We evaluate our approach on three challenging tasks: sequence and structure modeling, antigen-binding CDR design, and binding affinity optimization, and demonstrate superior performance compared to state-of-the-art methods, indicating the effectiveness of our model.

\section{Related Work}

\paragraph{Protein Design} 

Several machine learning approaches in structure-based protein design use fragment-based and energy-based global features derived from protein structures~\cite{hu2022protein,chen2020engineering,wu2021self,kuhlman2019advances}. A seminal work is~\cite{ingraham2019generative}'s introduction of the formulation of fixed-backbone design as a structure-to-sequence translation problem. GVP~\cite{jinglearning} developed typical model architectures with translational and rotational equivariances. GCA~\cite{tan2022generative} utilizes global attention to learn geometric representations from residue interactions. AlphaDesign~\cite{gao2022alphadesign} has established a benchmark based on AlphaFold DB~\cite{varadi2022alphafold,jumper2021highly}. ESM-IF~\cite{hsu2022learning} has augmented training data by incorporating predicted structures from AlphaFold2~\cite{jumper2021highly}. ProteinMPNN~\cite{dauparas2022robust} employs expressive features with message-passing networks similar to those used in \cite{ingraham2019generative}'s model. PiFold~\cite{gao2023pifold} introduces additional features and generates sequences in a one-shot manner. We focus on antibody design, a specific type of protein design, creating antibodies that bind to a target antigen with high affinity. 

\paragraph{Antibody Design}

Early approaches to computational antibody design relied heavily on handcrafted and statistical energy function optimization, utilizing Monte Carlo simulation to iteratively update both sequences and structures~\cite{pantazes2010optcdr,lapidoth2015abdesign,adolf2018rosettaantibodydesign,warszawski2019optimizing,ruffolo2021deciphering}. However, these methods are often computationally expensive and may only reach a local energy minimum. As an alternative, deep generative models have become a more feasible option. In the early stages of deep generative antibody design, sequence-based methods~\cite{alley2019unified,saka2021antibody,shin2021protein,akbar2022silico} were introduced. RefineGNN~\cite{jiniterative} is the first deep generative model of CDR sequence-structure co-design. DiffAb~\cite{luoantigen} generates antibodies explicitly targeting specific antigen structures by utilizing diffusion models. CEM~\cite{fu2022antibody} designs a constrained manifold to characterize the geometry constraints of the CDR loops. MEAN~\cite{kong2023conditional} applies E(3)-equivariant message passing and attention mechanisms. Our model aims to enhance capturing geometrical correlations between antigens and antibodies by incorporating structural information through a protein complex invariant embedding. Additionally, we developed a model capable of generating both CDR sequences and structures in a one-shot manner.

\paragraph{Generative Models for Molecules}

Autoregressive models have gained popularity for generating graphs~\cite{CCGC,DealMVC,wu2022graphmixup} in the context of biological molecules, as evidenced by studies such as GraphRNN~\cite{you2018graphrnn}, LGP-Net~\cite{li2018learning}, CG-VAE~\cite{liu2018constrained}, and HierVAE~\cite{jin2020hierarchical}. Among these, G-SchNet~\cite{gebauer2019symmetry} generates edges sequentially, while Graphite~\cite{grover2019graphite} iteratively refines the adjacency matrix with given node labels. RefineGNN~\cite{jiniterative} combines autoregressive models with iterative refinement to generate complete graphs with both node and edge labels for antibody design. MEAN~\cite{kong2023conditional} utilizes the multi-channel extension~\cite{huangequivariant} of the E(n)-equivariant GNN~\cite{satorras2021n} to generate sequences and structures in a progressive full-shot decoding manner. Recent advances in diffusion models~\cite{song2019generative,ho2020denoising,cao2022survey} have motivated their development in molecular generation, and several works~\cite{wu2022protein,trippe2022diffusion,CONVERT,gao2023diffsds} have successfully employed generative diffusion models in protein structure generation. In particular, DiffAb~\cite{luoantigen} has designed a diffusion probabilistic model specifically for antibody structure generation. While previous works have utilized either equivariant graph neural networks with iterative refinement or diffusion models with iterative sampling, we propose cross-gate MLPs that effectively capture geometric correlations and generate both CDR sequences and structures in a one-shot manner.

\section{Method}

\subsection{Preliminaries}

A protein complex is comprised of $N$ amino acids, which can be represented as characters in a sequence, denoted as $\mathcal{S} = \{s_i\}_{i=1}^N$. Each token $s_i$ in the sequence is referred to as a $\textit{residue}$, with a value that can be any one of the 20 amino acids. The three-dimensional structure of the protein is represented by the backbone atom coordinates, denoted as $\mathcal{X} = \{ \boldsymbol{x}_{i, \omega}\}_{i=1}^{N}$, where $\boldsymbol{x}_{i, \omega}\in\mathbb{R}^3$ and $\omega \in \{\mathrm{C\alpha, N, C, O}\}$. The antibody-antigen complex, a common type of protein complex, can be represented by the pair $\mathcal{C} = (\mathcal{S}, \mathcal{X})$.

\begin{figure}[ht]
  \centering
  \includegraphics[width=1.0\linewidth]{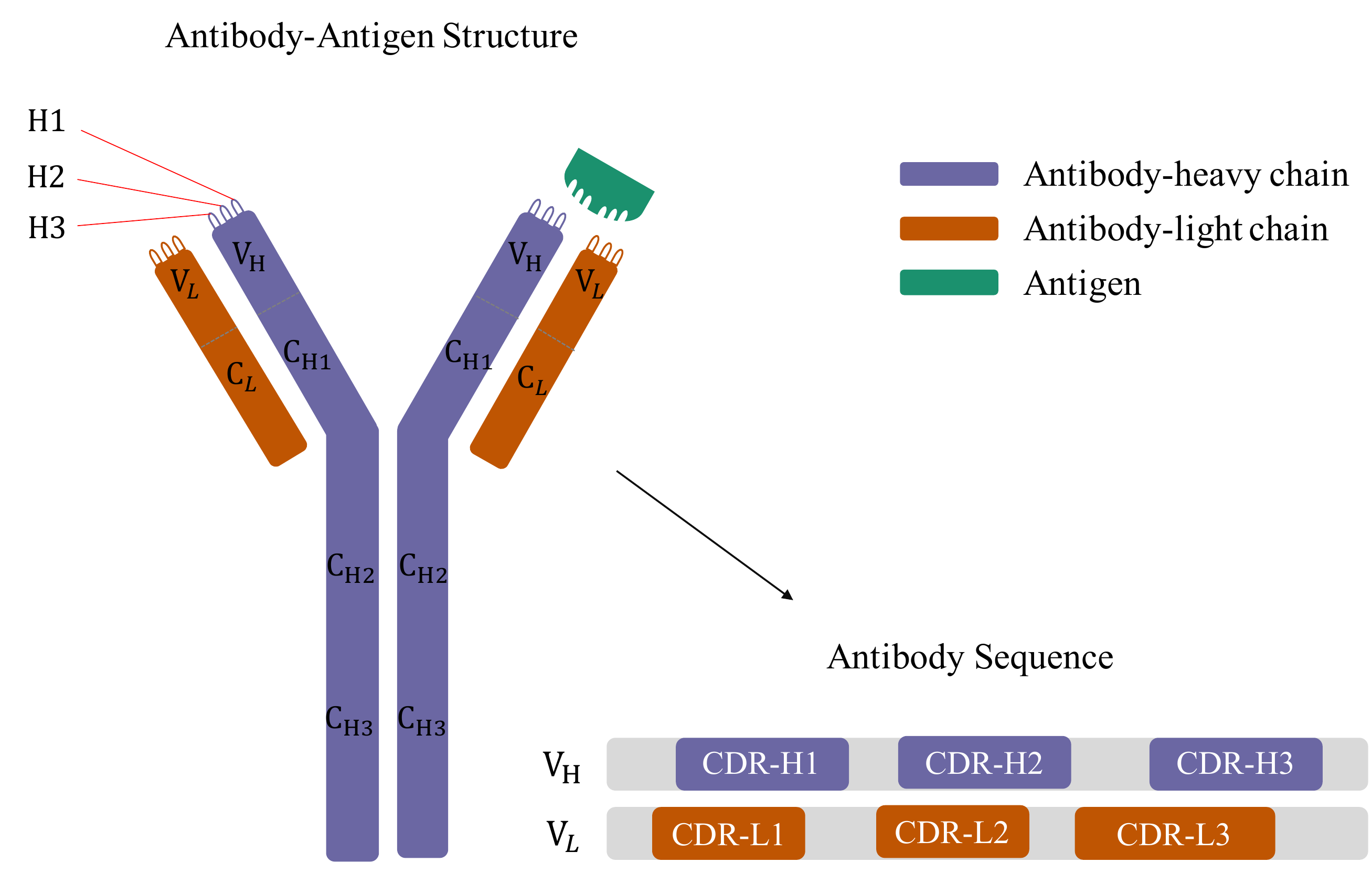}
  \caption{The schematic diagram of an antibody-antigen complex structure. Note that the antibody is a symmetric Y shape, each half of which contains a heavy and light chain. Here we focus on designing the CDR-H1, CDR-H2, and CDR-H3 loops in the heavy chain.}
  \label{fig:problem}
\end{figure}

Specifically, an antibody is a protein with a symmetrical Y shape, as depicted in Figure~\ref{fig:problem}. It consists of two identical H/L chains, each of which contains a \textit{variable domain} (VH/VL) and several constant domains. The variable domain can be further divided into a \textit{framework region}~\cite{jiniterative,kuroda2012computer} and three \textit{complementarity-determining regions} (CDRs), which play a crucial role in binding affinity to specific antigens. The heavy and light chains on each half of the antibody contain six CDR loops, namely CDR-H1, CDR-H2, CDR-H3, CDR-L1, CDR-L2, and CDR-L3, respectively. Prior research has formalized the antibody design problem as identifying CDRs that fit within a given framework region~\cite{shin2021protein,akbar2022silico}. More recent studies have indicated that CDRs in heavy chains are the most influential in determining antigen-binding affinity and are therefore the most difficult to design~\cite{jiniterative,fischman2018computational}. As a result, the context of the antigen and light chains are incorporated for better controlling the binding specificity of the generated antibodies.

The CDR is represented as a set of residues $\mathcal{R}$ given by:
\begin{equation}
  \mathcal{R}:=\big\{(s_i, \boldsymbol{x}_{i, \omega})|i=\{p+1,...,p+q\}\big\},
\end{equation}
The remainder context is represented as:
\begin{equation}
\mathcal{C}\setminus\mathcal{R}:=\big\{(s_i, \boldsymbol{x}_{i, \omega})|i=\{1,...,N\}\setminus\{p+1,...,p+q\}\big\}.
\end{equation}
where $\mathcal{C}$ is the full antibody-antigen complex, and $\mathcal{C}\setminus\mathcal{R}$ is the context information that excludes the CDR residues. Our objective is to learn a mapping $\mathcal{F}_\Theta: \mathcal{C}\setminus\mathcal{R} \mapsto \mathcal{C}$, parameterized by $\Theta$, that generates the sequence and structure of a CDR consisting of $q$ amino acids with indices $p+1$ to $p+q$. The optimal parameters $\Theta^*$ of $\mathcal{F}_\Theta$ is obtained by:
\begin{equation}
  \Theta^* = \arg\min_\Theta \mathcal{L}(\mathcal{F}_\Theta(\mathcal{C}\setminus\mathcal{R}), \mathcal{C}),
\end{equation}
where $\mathcal{L}$ is a loss function that measures the difference between the generated and real antibody-antigen complex.

\subsection{Overview}
\label{sec:overview}
\begin{figure}[ht]
  \centering
  \includegraphics[width=0.98\linewidth]{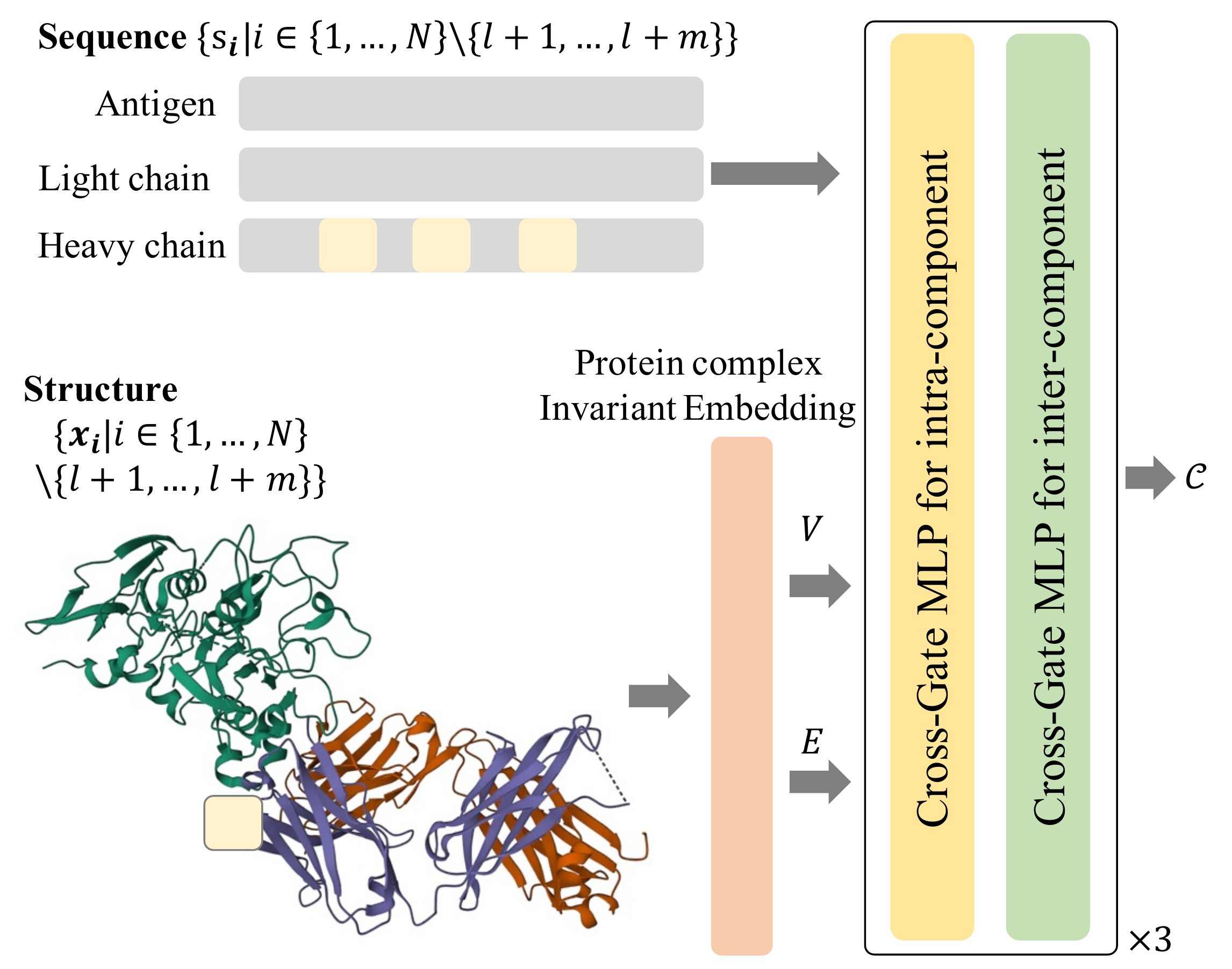}
  \caption{The overall framework of our model. The input is both the sequence and structure of the antibody-antigen complex. The CDRs are visually masked by light yellow blocks to highlight their generation by the model. The output consists of the complete sequence and structure of the antibody-antigen complex, including the generated CDRs.}
  \label{fig:overview}
\end{figure}

The overall framework is depicted in Figure~\ref{fig:overview}, named as \textbf{ADesigner}. The input to the model is the structure of the antibody-antigen complex, which is processed by a protein complex invariant embedding (PIE) module to obtain two sets of embeddings: one for intra-component interactions and the other for inter-component interactions. The PIE module is explained in detail in the following subsection.

The embeddings undergo processing through a sequence of cross-gate MLP modules, which take into account both the sequence and structure information. In each block of cross-gate MLP modules, there are two types of modules for processing intra- and inter-component interactions separately. The cross-gate MLP modules facilitate sequence-structure co-learning, allowing the representations to be refined and enriched in an implicit manner.

Ultimately, the learned embeddings are harnessed to generate the complete sequence and structure of the antibody-antigen complex. In contrast to prior methods that depend on explicit iterative decoding strategies, our framework directly outputs the generated result. This is made possible by the co-learning of sequence and structure information in the cross-gate MLP modules. Overall, our framework provides a more efficient and effective approach to generating the antibody-antigen complex.

\subsection{Protein Complex Invariant Embedding}
\label{sec:pcie}

In order to obtain a comprehensive geometric model of the antibody-antigen complex, we introduce the concept of the \textit{Protein complex Invariant Embedding} (PIE). Following previous works~\cite{jiniterative,kong2023conditional}, we represent the protein structure as a graph, where $S_i$ denotes the component of residue $i$. We define intra-component and inter-component edges as follows:

\begin{figure}[ht]
\centering
\includegraphics[width=1.0\linewidth]{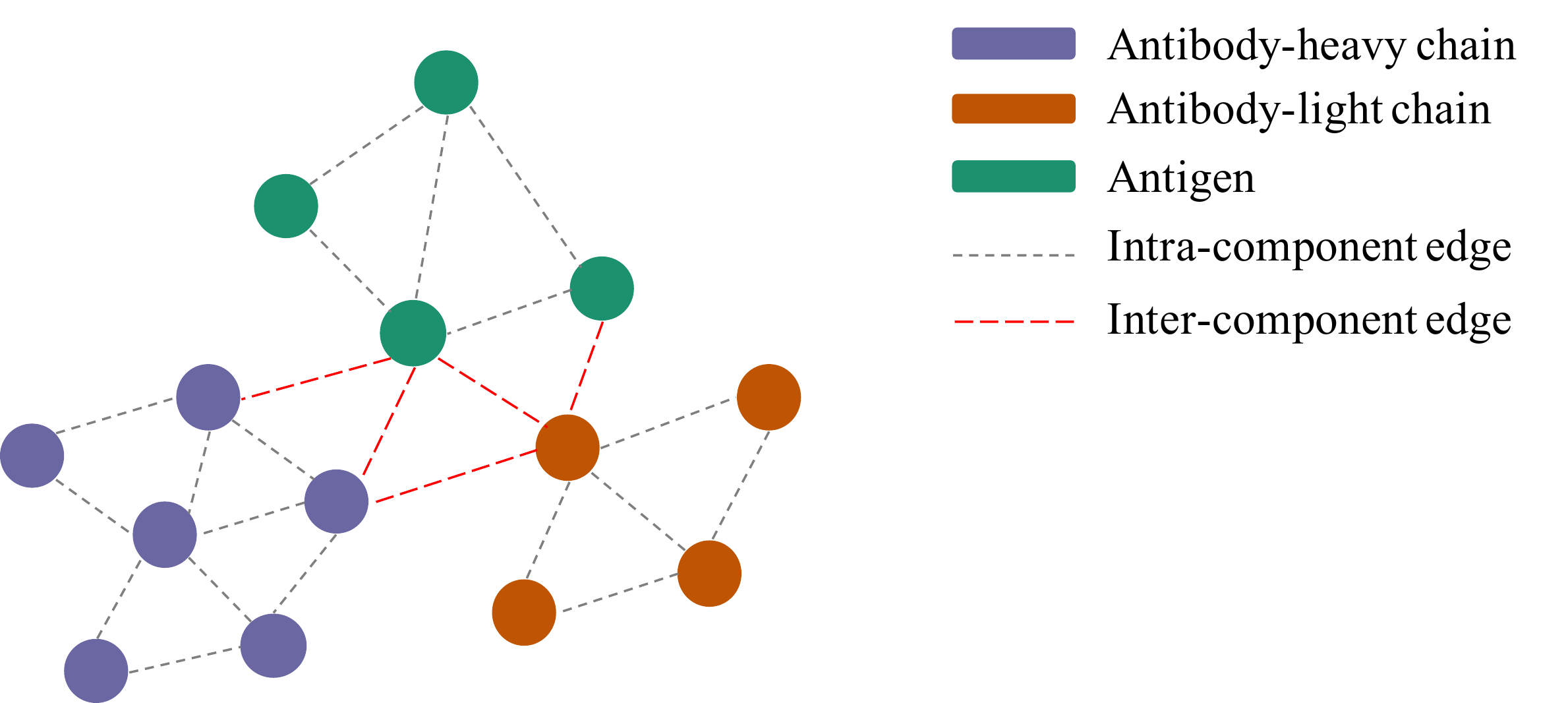}
\caption{The schematic diagram of intra- and inter-component edges in the antibody-antigen complex.}
\label{fig:edge}
\end{figure}

\textbf{Definition 1.} An \textit{intra-component edge} is defined as the edge between two residues in the same component if the distance between their C$\alpha$ atoms is less than a threshold $\delta_{in}$. For residue $i$, we denote the set of its intra-component edges as $\mathcal{E}_{in}(i) = \{j \; \vert \; \|\boldsymbol{x}_{i, \mathrm{C}\alpha} - \boldsymbol{x}_{j, \mathrm{C}\alpha}\|^2 < \delta_{in}, \forall S_i = S_j \}$.

\textbf{Definition 2.} An \textit{inter-component edge} is defined as the edge between two residues in different components if the distance between their C$\alpha$ atoms is less than a threshold $\delta_{ex}$. For residue $i$, we denote the set of its inter-component edges as $\mathcal{E}_{ex}(i) = \{j \; \vert \; \|\boldsymbol{x}_{i, \mathrm{C}\alpha} - \boldsymbol{x}_{j, \mathrm{C}\alpha}\|^2 < \delta_{ex}, \forall S_i \neq S_j\}$.

The intra- and inter-component edges are defined for general protein complexes. In the case of our antibody-antigen complex, there are three components: the heavy chain, the light chain, and the antigen. The schematic diagram of intra- and inter-component edges is shown in Figure~\ref{fig:edge}. Empirically, we set the thresholds $\delta_{in} = 8.0\textup{\AA}$ and $\delta_{ex} = 12.0\textup{\AA}$.

Given that the intra- and inter-component edges have captured the component-level interactions, our focus now turns to the residue-level dependencies. We achieve this by transforming the protein complex structure coordinates into a graph. For each residue $i$, we define its node embedding as the distance encoding of its C$\alpha$ atom to the remaining backbone atoms:
\begin{equation}
  V_i = \bigg\{\mathrm{RBF}(\|\boldsymbol{x}_{i, \omega} - \boldsymbol{x}_{i, \gamma}\|) \; \bigg| \; \omega, \gamma \in \{\mathrm{Ca, N, C, O}\} \bigg\},
\end{equation}
where $\mathrm{RBF}(\cdot)$ is a radial basis distance encoding function. Analogously, we define edge embedding as the distance encoding of pairwise backbone atoms in the neighboring residue. We also encode the directions to identify the relative positions between neighboring residues. Formally, the intra- and inter-component edge embeddings are as follows:
\begin{equation}
\begin{aligned}
E_i^{in} = \bigg\{&\mathrm{RBF}(\|\boldsymbol{x}_{i, \omega} - \boldsymbol{x}_{j, \gamma}\|), \; \boldsymbol{Q}_i^T \frac{\boldsymbol{x}_{i, \omega} - \boldsymbol{x}_{j, \gamma}}{\|\boldsymbol{x}_{i, \omega} - \boldsymbol{x}_{j, \gamma}\|} \; \bigg| \; \\ &\omega, \gamma \in \{\mathrm{Ca, N, C, O}\}, j \in \mathcal{E}_{in} \bigg\},\\
E_i^{out} = \bigg\{&\mathrm{RBF}(\|\boldsymbol{x}_{i, \omega} - \boldsymbol{x}_{j, \gamma}\|), \; \boldsymbol{Q}_i^T \frac{\boldsymbol{x}_{i, \omega} - \boldsymbol{x}_{j, \gamma}}{\|\boldsymbol{x}_{i, \omega} - \boldsymbol{x}_{j, \gamma}\|} \; \bigg| \; \\ &\omega, \gamma \in \{\mathrm{Ca, N, C, O}\}, j \in \mathcal{E}_{out} \bigg\},
\end{aligned}
\end{equation}
where $\boldsymbol{Q}_i$ is a local coordinate system~\cite{ingraham2019generative} of residue $i$.

\subsection{Cross-Gate MLP}

To improve the efficiency of protein complex sequence-structure co-learning, we propose a novel Cross-Gate MLP (CGMLP) that updates sequence embeddings by incorporating both sequence and structure embeddings.

\begin{figure}[ht]
  \centering
  \includegraphics[width=0.98\linewidth]{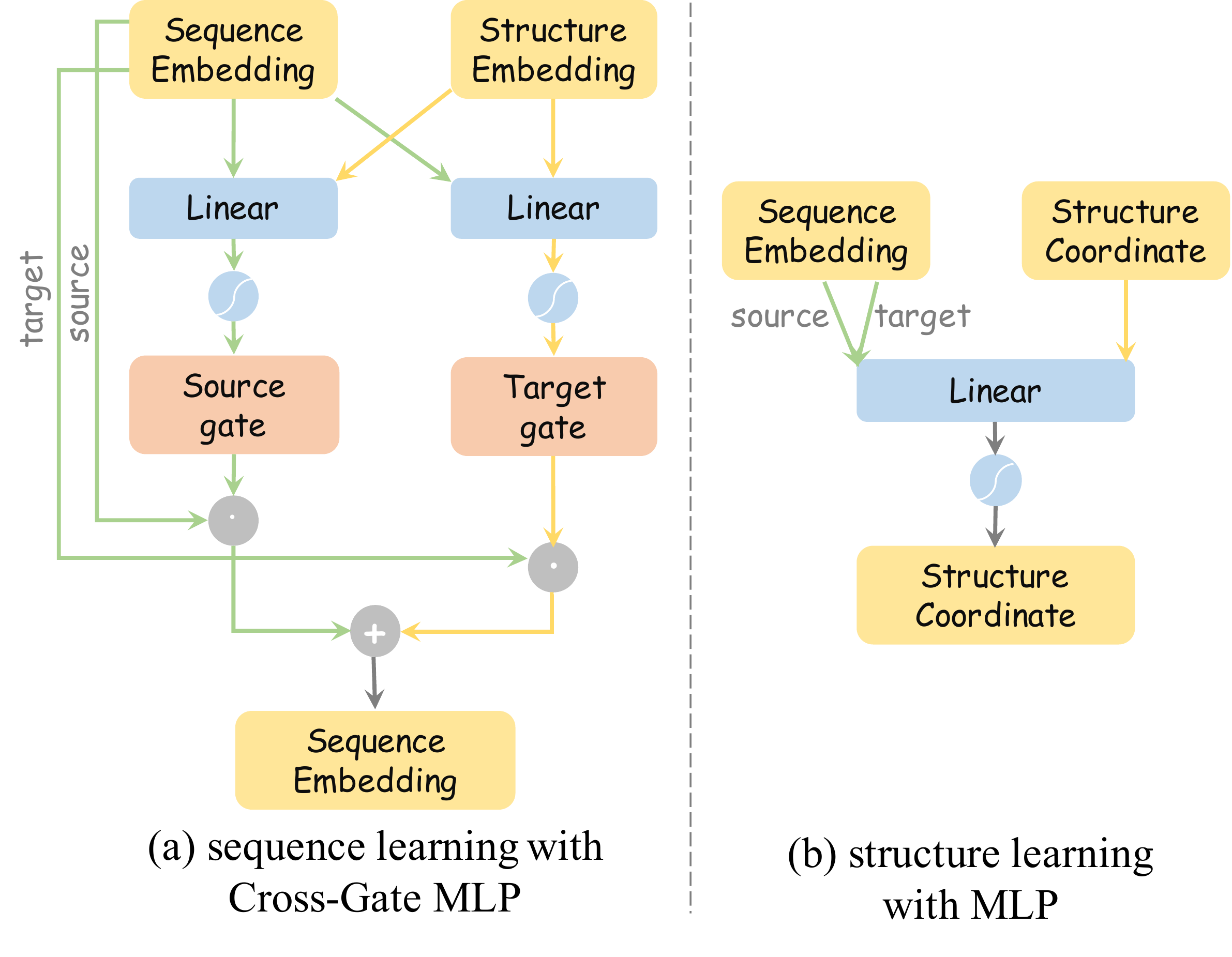}
  \caption{The schematic diagram of the sequence learning with Cross-Gate MLP and structure learning with MLP.}
  \label{fig:crossmlp}
\end{figure}

We define $\boldsymbol{h}_i^{(l)}$ as the source sequence embedding of residue $i$ in layer $l$, and its target sequence embedding $\boldsymbol{h}_j^{(l)}$ in layer $l$ is from its neighboring residue $j$. The coordinates of residue $i$ in layer $l$ are denoted as $\boldsymbol{Z}_i^{(l)} \in \mathbb{R}^{4\times 3}$, which contains the four types of backbone atoms $\{\boldsymbol{x}_{i, \omega} | \omega \in \{Ca, N, C, O\}\}$. The CGMLP is defined as follows:
\begin{equation}
  m_{ij}^{(l)} = \mathrm{Concat}(\boldsymbol{h}_i^{(l)}, \boldsymbol{h}_j^{(l)}, V_i, V_j, E_i),
  \label{equ:mij}
\end{equation}
\begin{equation}
  g_{s}^{(l)} = \sigma(\phi_s(m_{ij}^{(l)})), g_{t}^{(l)} = \sigma(\phi_t(m_{ij}^{(l)})),
\end{equation}
\begin{equation}
  m_{ij}^{(l+1)} = g_{s}^{(l)} \odot \boldsymbol{h}_j^{(l)} + g_{t}^{(l)} \odot \boldsymbol{h}_i^{(l)},
\end{equation}
\begin{equation}
  \boldsymbol{h}_i^{(l+1)} = \phi_h(\boldsymbol{h}^{(l)}, \sum_{j \in \mathcal{E}} m_{ij}^{(l+1)}),
\end{equation}
where $\phi_s(\cdot)$, $\phi_t(\cdot)$, and $\phi_h(\cdot)$ are MLPs, $\sigma(\cdot)$ is the Sigmoid activation function, $\odot$ is the element-wise multiplication, and $\mathrm{Concat}(\cdot)$ is the concatenation operation. Here, we use $\mathcal{E}$ and $E_i$ without the subscript to denote both the set of \textit{intra- and inter-component edges}, and \textit{the edge embedding of residue $i$}, respectively, for convenience.

As illustrated in Figure~\ref{fig:crossmlp}(a), we utilize both the sequence and structure embeddings to obtain the latent message $m_{ij}^{l}$. The source and target gates are obtained by using $m_{ij}^{l}$ with two individual branches of an MLP and a sigmoid activation function. The refined latent message $m_{ij}^{l+1}$ is the sum of the source and target sequence embeddings weighted by the gates. Finally, the sequence embedding $\boldsymbol{h}_i^{(l+1)}$ of residue $i$ in layer $l+1$ is obtained by aggregating the refined latent message of its neighboring residues.

With the refined sequence embedding, we update the coordinates $\boldsymbol{Z}_i^{(l+1)}$ of residue $i$ in layer $l+1$ as follows:
\begin{equation}
  m_{ij}'^{(l)} = \mathrm{Concat}(\boldsymbol{h}_i^{(l)}, \boldsymbol{h}_j^{(l)}, \frac{(\boldsymbol{Z}_{i}^{(l)} - \boldsymbol{Z}_{j}^{(l)})^T (\boldsymbol{Z}_{i}^{(l)} - \boldsymbol{Z}_{j}^{(l)})}{\|(\boldsymbol{Z}_{i}^{(l)} - \boldsymbol{Z}_{j}^{(l)})^T (\boldsymbol{Z}_{i}^{(l)} - \boldsymbol{Z}_{j}^{(l)})\|_F}),
\end{equation}

\begin{equation}
  \boldsymbol{Z}^{(l+1)} = Z^{(l)} + \frac{1}{|\mathcal{E}|} \sum_{j\in \mathcal{E}} \phi_z(m_{ij}'^{(l)}) (\boldsymbol{Z}_{i}^{(l)} - \boldsymbol{Z}_{j}^{(l)}),
\end{equation}
where, $\phi_z(\cdot)$ is a vanilla MLP. As illustrated in Figure~\ref{fig:crossmlp}(b), the coordinates of residue $i$ in layer $l+1$ are obtained by aggregating the refined latent messages from its neighboring residues. It's worth noting that we only use refined sequence embeddings to update the structure coordinates, rather than structure embeddings. The reason is that structure embeddings are directly influenced by the coordinates, and using them may lead to overfitting of structure learning.

\subsection{One-shot Decoding}

Our method has been designed to perform sequence-structure co-learning, allowing us to directly output the CDR regions of the antibody-antigen complex without the need for any additional decoding process. Assuming there are $L$ layers, the predicted sequence $\hat{s}_i$ and structure $\hat{\boldsymbol{Z}}_i$ of the CDR regions are obtained as follows:
\begin{equation}
\begin{aligned}
  \hat{s}_i &= \mathrm{Argmax}(h_i^{(L)}), \\ \hat{\boldsymbol{Z}}_i&=\boldsymbol{Z}^{(L)}_i,
\end{aligned}
\end{equation}
where $\mathrm{Argmax}(\cdot)$ is the argmax operation.

The loss function is defined as a linear combination of the sequence loss and the structure loss. For the sequence loss, we use the cross-entropy loss between the predicted sequence and the ground truth sequence:
\begin{equation}
  \mathcal{L}_{seq} = \frac{1}{q} \sum_{i=p+1}^{p+q} \ell_{ce}(s_i, \mathrm{Softmax}(h_i^{(L)})),
\end{equation}
where $\ell_{ce}$ denotes the cross-entropy loss, and $\mathrm{Softmax}(\cdot)$ is the softmax activation function.
For the structure loss, we use the differentiable L1 loss~\cite{lai2018fast}:
\begin{equation}
  \mathcal{L}_{struct} = \frac{1}{q} \sum_{i=p+1}^{p+q} \sqrt{(\boldsymbol{Z}_i - \hat{\boldsymbol{Z}}_i)^2 + \epsilon^2},
\end{equation}
where $\epsilon$ is a small constant empirically set to $10^{-8}$. This loss function is more robust to outliers compared to the commonly used L2 loss, as it suppresses large errors using the $\epsilon^2$ term. Consequently, outliers do not have much influence on the total loss, making the network more stable. The overall loss is $\mathcal{L} = \mathcal{L}_{seq} + \lambda \mathcal{L}_{struct}$, where $\lambda = 0.8$ is a weight hyperparameter that balances the sequence and structure loss.

\section{Experiments}

We evaluate our model on three challenging antibody design tasks using the common experimental setups from previous works~\cite{jiniterative,kong2023conditional,fu2022antibody}. These tasks include: (i) generative task on the Structural Antibody Database~\cite{dunbar2014sabdab}, (ii) antigen-binding CDR-H3 design using an existing antibody design benchmark of 60 antibody-antigen complexes from~\cite{adolf2018rosettaantibodydesign}, and (iii) antigen-antibody binding affinity optimization that redesigns CDR-H3 of antibodies on the Structural Kinetic and Energetic database of Mutant Protein Interactions~\cite{jankauskaite2019skempi}. 

\paragraph{Baselines} We compare our model to recent state-of-the-art approaches, including (i) \textbf{LSTM}-based approach by~\cite{saka2021antibody,akbar2022silico} that generates the amino acid sequence in an autoregressive manner without structure modeling, leveraging a long short-term memory (LSTM) network; (ii) \textbf{C-LSTM} implemented by~\cite{kong2023conditional} that considers the entire context of the antibody-antigen complex, built upon LSTM; (iii) \textbf{RefineGNN} proposed by~\cite{jiniterative} that takes the 3D geometry for antibody CDR design. This approach unravels the amino acid sequence in an autoregressive manner and iteratively refines its predicted global structure. (iv) \textbf{C-RefineGNN} implemented by~\cite{kong2023conditional} that extends RefineGNN by accommodating the entire antibody-antigen complex. (v) \textbf{MEAN} proposed by~\cite{kong2023conditional}, which is related to but distinct from our method. It takes less geometric information into account and requires an iterative refinement strategy. We used the default setup of each method, training the models for 20 epochs with Adam optimizer and a learning rate of $10^{-3}$. We used the checkpoint with the lowest validation loss for testing.

\paragraph{Metrics} We evaluate the results from two perspectives, i.e., sequence modeling, and structure modeling. For sequence modeling, we employ Amino Acid Recovery (AAR) that measures the overlapping rate between the predicted sequences and ground truths. For structure modeling, we employ Root Mean Squared Deviation (RMSD) between the predicted structures and ground truths. We report the TM-score~\cite{zhang2004scoring,xu2010significant} that measures the global structural similarity in the second task.

\subsection{Sequence and Structure Modeling}

\begin{table*}[ht]
\label{tab:kfold}
\centering
\setlength{\tabcolsep}{3.8mm}{
\begin{tabular}{ccccccc}
\toprule
\multirow{2}{*}{Method} & \multicolumn{2}{c}{CDR-H1} & \multicolumn{2}{c}{CDR-H2} & \multicolumn{2}{c}{CDR-H3} \\
& AAR ($\uparrow$) & RMSD ($\downarrow$) & AAR ($\uparrow$) & RMSD($\downarrow$) & AAR ($\uparrow$) & RMSD($\downarrow$) \\
\midrule
LSTM & 49.98$\pm$5.20\% & - & 28.50$\pm$1.55\% & - & 15.69$\pm$0.91\% & - \\ 
C-LSTM & 40.93$\pm$5.41\% & - & 29.24$\pm$1.08\% & - & 15.48$\pm$1.17\% & - \\
RefineGNN               & 39.40$\pm$5.56\%                           & 3.22$\pm$0.29       & 37.06$\pm$3.09\%                            & 3.64$\pm$0.40                           & 21.13$\pm$1.59\%                           & 6.00$\pm$0.55                           \\
C-RefineGNN             & 33.19$\pm$2.99\%                           & 3.25$\pm$0.40       & 33.53$\pm$3.23\%                           & 3.69$\pm$0.56                           & 18.88$\pm$1.37\%                           & 6.22$\pm$0.59                           \\
DiffAB                    & 61.34$\pm$1.98\%            &   1.02$\pm$0.66     & 37.66$\pm$1.89\%                      & 1.20$\pm$0.09            & 25.79$\pm$1.52\%                      & 3.02$\pm$0.11                         \\
MEAN                    & 58.29$\pm$7.26\%                           & 0.98$\pm$0.16       & 47.15$\pm$3.09\%                           & 0.95$\pm$0.05                           & 36.38$\pm$3.08\%                           & 2.21$\pm$0.16                           \\
ADesigner                    & \textbf{64.34}$\pm$3.37\% & \textbf{0.82}$\pm$0.12       & \textbf{55.52}$\pm$3.36\% & \textbf{0.79}$\pm$0.06 & \textbf{37.37}$\pm$2.33\% & \textbf{1.97}$\pm$0.19 \\
\hline
Improvement & +6.05\% & +16.33\% & +8.37\% & +16.84\% & +0.98\% & +10.86\% \\
\bottomrule
\end{tabular}}
\caption{The mean (standard deviation) of 10-fold cross-validation results for 1D sequence and 3D structure modeling on the SAbDab dataset.}
\end{table*}

We evaluate our model with baseline approaches on the Structural Antibody Database (SAbDab)~\cite{dunbar2014sabdab}, which contains 3,127 complexes consisting of heavy chains, light chains, and antigens. Following~\cite{jiniterative,kong2023conditional}, the dataset is split into training, validation, and testing sets according to the clustering of CDRs to maintain the generalization test. The total numbers of clusters for CDR-H1, CDR-H2, and CDR-H3 are 765, 1093, and 1659, respectively. The clusters are split into training, validation, and testing sets with a ratio of 8:1:1. We report the results of 10-fold cross-validation in Table 1.

It can be observed that our proposed method surpasses all other methods in terms of AAR and RMSD scores for all three CDR regions. Furthermore, our proposed method outperforms MEAN by a significant margin, with an average improvement of over 5.13\% in AAR and over 14.68\% in RMSD. These results demonstrate the efficacy of our proposed approach in modeling both the sequence and structure of CDRs, making it a promising method for the sequence and structure modeling of antibody-antigen complexes.

\subsection{Antigen-Binding CDR-H3 Design}

We validated our approach for designing CDR-H3 loops with desired antigen-binding capabilities using the well-established RAbD benchmark dataset~\cite{adolf2018rosettaantibodydesign}. For a comprehensive comparison, the widely-adopted conventional method, RosettaAD~\cite{adolf2018rosettaantibodydesign}, is also incorporated as a benchmark. Rigorous training was undertaken using the extensive SAbDab database of antibody-antigen complexes, carefully excluding any entries bearing significant structural homology to complexes present in the RAbD test set. A detailed analysis of the results is provided in Table 2.

\begin{table}[ht]
\label{tab:rabd}
\centering
\setlength{\tabcolsep}{2mm}{
\begin{tabular}{cccc}
\toprule
\multirow{2}{*}{Method} & \multicolumn{3}{c}{CDR-H3}                                                                            \\
                        & AAR ($\uparrow$)                  & TM-score ($\uparrow$)            & RMSD ($\downarrow$)            \\
\midrule
RosettaAD               & 22.50\%                           & 0.9435                           & 5.52                           \\
LSTM & 22.36\% & - & - \\
C-LSTM & 22.18\% & - & - \\
RefineGNN               & 29.79\%                           & 0.8308                           & 7.55                           \\
C-RefineGNN             & 28.90\%                           & 0.8317                           & 7.21                           \\
MEAN                    & 36.77\%                           & 0.9812                           & 1.81                           \\
\hline
ADesigner                    & \textbf{40.94}\% & \textbf{0.9850} & \textbf{1.55} \\
\bottomrule
\end{tabular}}
\caption{The performance of CDR-H3 design on the RAbD benchmark using amino acid recovery (AAR), TM-score, and RMSD metrics.}
\end{table}

The results clearly demonstrate the superior performance of our method compared to all other techniques, as evidenced by the improved accuracy across AAR, TM-score, and RMSD metrics. This highlights the efficacy of our approach for designing CDR-H3 loops that closely recapitulate native antigen-binding topologies. 

\begin{figure}[ht]
\centering
\includegraphics[width=1.0\linewidth]{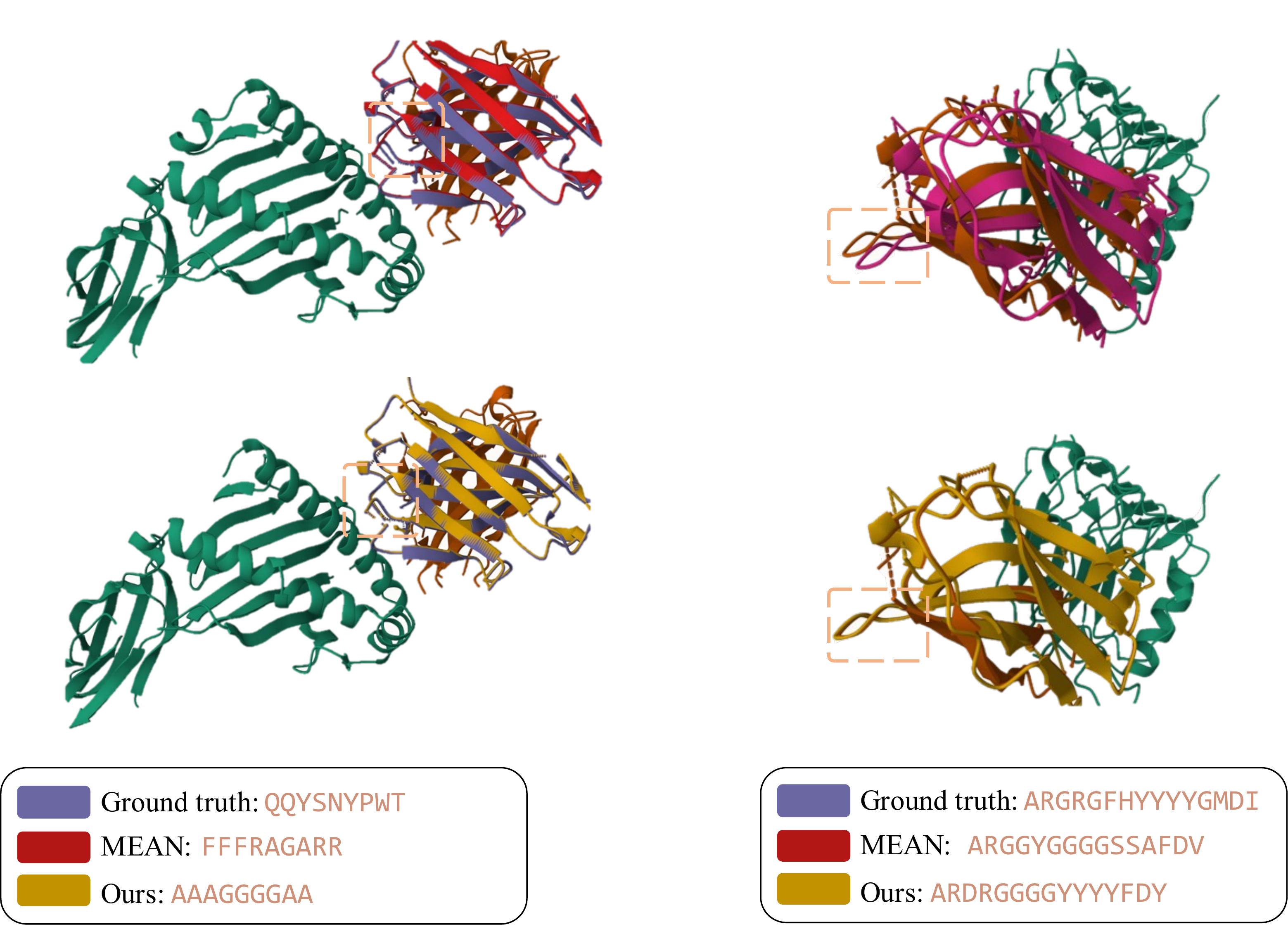}
\vspace{2mm}
\caption{The designed examples of CDR-H3.}
\label{fig:designed_example}
\end{figure}

To provide further insights, we visually examine the designed CDR-H3 structures for two representative antibody test cases (PDB IDs: 1w72 and 3h3b) in Figure~\ref{fig:designed_example}. For the first example, the loops designed by MEAN and our method yield RMSDs of $0.27\textup{\AA}$ and $0.19\textup{\AA}$ respectively, compared to the native structure. While MEAN performs reasonably well, it still contains inaccuracies in some details. In contrast, our approach predicts a structure that closely matches the native one. In the second more challenging example, the CDR-H3 loops designed by MEAN and our approach have RMSDs of $2.46\textup{\AA}$ and $1.19\textup{\AA}$ respectively. MEAN fails on this difficult sample, whereas our method predicts a structure that approximates the correct native conformation.

\begin{table*}[ht]
\centering
\setlength{\tabcolsep}{4mm}{
\begin{tabular}{ccccccc}
\toprule
\multirow{2}{*}{Method} & \multicolumn{2}{c}{CDR-H1} & \multicolumn{2}{c}{CDR-H2} & \multicolumn{2}{c}{CDR-H3} \\
& AAR ($\uparrow$) & RMSD ($\downarrow$) & AAR ($\uparrow$) & RMSD($\downarrow$) & AAR ($\uparrow$) & RMSD($\downarrow$) \\
\midrule
ADesigner & 64.34$\pm$3.37\% & 0.82$\pm$0.12 & 55.52$\pm$3.36\% & 0.79$\pm$0.06 & 37.37$\pm$2.33\% & 1.97$\pm$0.19 \\
\hline
w/o PIE & 60.27$\pm$6.69\% & 0.95$\pm$0.16 & 49.14$\pm$2.96\% & 0.98$\pm$0.23 & 35.78$\pm$2.43\% & 2.17$\pm$0.21 \\
w/o CGMLP & 62.59$\pm$5.09\% & 1.07$\pm$0.17 & 52.50$\pm$3.78\% & 0.97$\pm$0.18 & 36.14$\pm$2.56\% & 2.00$\pm$0.21 \\
\bottomrule
\end{tabular}}
\label{tab:ablation}
\caption{Ablation of our proposed method on the SAbDab dataset.}
\end{table*}

\subsection{Affinity Optimization}

We thoroughly evaluated the efficacy of our methodology for optimizing the binding affinity of antibody-antigen complexes through the simultaneous sequence and conformational tuning of the crucial CDR-H3 loop region. To predict the binding energy ($\Delta\Delta \mathrm{G}$) after optimization, we utilized the pre-trained deep geometric network~\cite{shan2022deep} and followed the same protocol as in a previous study~\cite{kong2023conditional}. We incorporated Iterative Target Augmentation~\cite{yang2020improving} (ITA) into the optimization process. The results are presented in Table 4.

\begin{table}[ht]
\centering
\setlength{\tabcolsep}{12mm}{
\begin{tabular}{cc}
\toprule
Method      & $\Delta\Delta$G ($\downarrow$)   \\
\midrule
Random      & +1.52                            \\
LSTM        & -1.48                            \\
C-LSTM      & -1.83                            \\
RefineGNN   & -3.98                            \\
C-RefineGNN & -3.79                            \\
MEAN        & -5.33                            \\
\hline
ADesigner        & -\textbf{10.78} \\
\bottomrule
\end{tabular}}  
\label{tab:aff_opt}
\caption{Average affinity change after optimization. The lower is better.}
\end{table}

\begin{figure}[ht]
  \centering
  \includegraphics[width=1.0\linewidth]{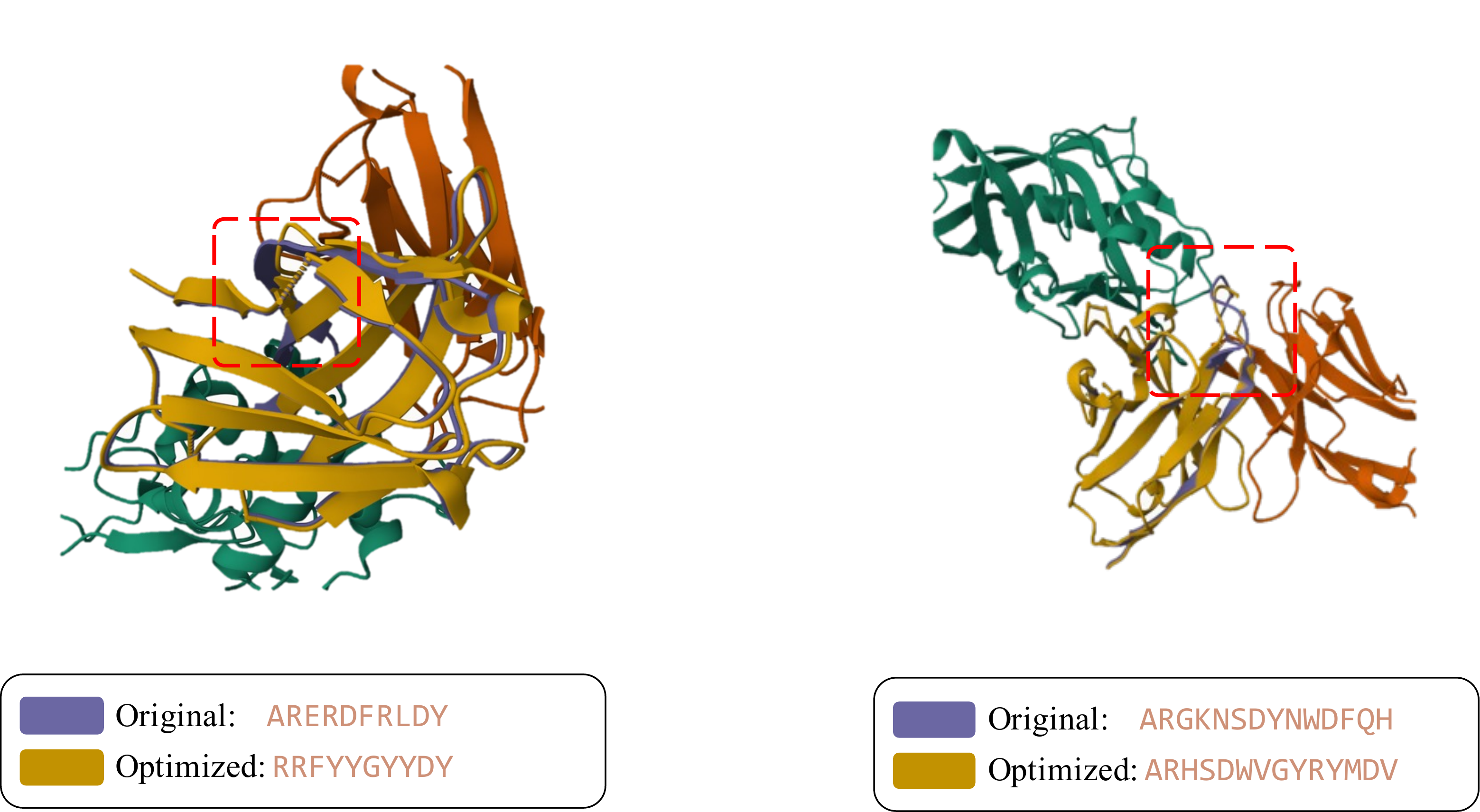}
  \caption{The optimized examples of CDR-H3.}
  \label{fig:aff_opt_vis}
\end{figure}

Our proposed method achieved superior results to the previous state-of-the-art method MEAN, with a notably lower $\Delta\Delta \mathrm{G}$ of below $-10 \mathrm{kcal/mol}$, indicating a substantial binding affinity between the optimized antibody and the antigen. We visualize two optimized examples (PDB IDs: 1kip, $\Delta\Delta G=-10.60$; 4jpk, $\Delta\Delta G=-12.09$) in Figure~\ref{fig:aff_opt_vis}.

\subsection{Ablation Study}

We conducted ablation studies as summarized in Table 3. Specifically, we examined the impact of removing the protein complex invariant embedding (w/o PIE) and replacing the cross-gate MLP with equivariant graph neural networks~\cite{satorras2021n,kong2023conditional} (w/o CGMLP). Our results show that PIE provides rich information that plays a critical role in our model. Furthermore, the CGMLP consistently improved the performance across AAR and RMSD metrics. These findings demonstrate the effectiveness of our approach.

\subsection{Training/Inference Efficiency} 
\label{app:efficiency}

We conducted a comparison of the training and inference efficiency. Training efficiency was the training time of one full epoch on the SAbDab training set, while inference efficiency was the inference time on the SAbDab testing set, including postprocessing steps that may result in more time overhead. Table 5 shows that our method outperforms RefineGNN and MEAN in both training and inference efficiency. The training efficiency of our method is 16s, which is 27.27\% faster than MEAN. Meanwhile, our method's inference efficiency is 24s, which is 14.29\% faster than RefineGNN and 14.29\% faster than MEAN. These results demonstrate that our one-shot design is highly efficient and effective in optimizing antibody sequence and structure design.

\begin{table}[ht]
\small
\centering
\setlength{\tabcolsep}{1.5mm}{
\begin{tabular}{ccc}
\toprule
Method    & Training efficiency ($\downarrow$) & Inference efficiency ($\downarrow$) \\
\midrule
RefineGNN & 218s & 47s \\
MEAN      & 22s  & 28s \\
ADesigner      & 14s  & 24s \\
\hline
Improvement & +36.36\% & +14.29\% \\
\bottomrule     
\end{tabular}}
\label{tab:efficiency}
\caption{The training and inference efficiency comparison.}
\end{table}

\section{Conclusions and Limitations}

In this paper, we develop a simple yet effective antibody designer for antibody sequence and structure design based on the entire context of the antibody-antigen complex. By leveraging comprehensive geometric modeling with a novel macromolecular invariant embedding tailored for protein complexes, and enabling sequence-structure co-learning through a simple cross-gate MLP, our approach achieves competitive results on various antibody-related tasks. A limitation is that our method is currently limited to in silico design; we leave wet-lab validation to future work. 

\section{Acknowledgements}
This work was supported by the National Key R\&D Program of China (2022ZD0115100), the National Natural Science Foundation of China (U21A20427), the Competitive Research Fund (WU2022A009) from the Westlake Center for Synthetic Biology and Integrated Bioengineering.

\bibliography{ref}

\end{document}